\documentclass[jcs]{iosart2x}

\usepackage{gensymb}

\pubyear{0000}
\volume{0}
\firstpage{1}
\lastpage{1}

\begin{document}

\begin{frontmatter}

\textcolor{blue}{\textbf{\\ Disclaimer: }
© Md Morshed Alam, Weichao Wang, 2021. 
The definitive, peer reviewed and edited version of this article is published in Journal of Computer Security, volume 29, issue 4, pages 423-446, 2021, http://dx.doi.org/10.3233/JCS-200108}

\title{A Comprehensive Survey on the State-of-the-art Data Provenance Approaches for Security Enforcement}
\runtitle{A Comprehensive Survey on Data Provenance Approaches for Security Enforcement}


\author{\inits{M.M.}\fnms{Md Morshed} \snm{Alam}\ead[label=e1]{malam3@uncc.edu}%
\thanks{Corresponding author. \printead{e1}.}} 
and
\author{\inits{W.}\fnms{Weichao} \snm{Wang}\ead[label=e2]{wwang22@uncc.edu}}
\address{Department of Software and Information Systems, \orgname{University of North Carolina at Charlotte},
NC, \cny{USA}\printead[presep={\\}]{e1,e2}}

\begin{abstract}
Data provenance collects comprehensive information about the events and operations in a computer system at both application and system levels. It provides a detailed and accurate history of transactions that help delineate the data flow scenario across the whole system. Data provenance helps achieve system resilience by uncovering several malicious attack traces after a system compromise that are leveraged by the analyzer to understand the attack behavior and discover the level of damage. Existing literature demonstrates a number of research efforts on information capture, management, and analysis of data provenance. In recent years, provenance in IoT devices attracts several research efforts because of the proliferation of commodity IoT devices. In this survey paper, we present a comparative study of the state-of-the-art approaches to provenance by classifying them based on frameworks, deployed techniques, and subjects of interest. We also discuss the emergence and scope of data provenance in IoT network. Finally, we present the urgency in several directions that data provenance needs to pursue, including data management and analysis.

\end{abstract}

\begin{keyword}
\kwd{Data Provenance}
\kwd{Information Flow Control}
\kwd{Digital Forensic}
\kwd{Attack Reconstruction}
\kwd{Taint Propagation}
\end{keyword}

\end{frontmatter}


\section{Introduction}

\par Data provenance contains the complete history of operations on data and processes starting from the system boot-up. It provides sufficient details of data ownership changes, data manipulation, and process activities \cite{Bunemanetal2001} \cite{Simmhanetal2005} \cite{Zafar-Trust-17}. Data provenance explains how data evolves from process to process. It clearly depicts the relationship between inputs and outputs of a process, which is important to infer the characteristics of that process (and application that holds it). In the case of a cyber attack, it produces adequate system traces to regenerate a successful (or unsuccessful) attack for design of mitigation mechanisms. A \textit{trace} is basically an imprint of past events or activities (malicious or benign) in the system \cite{osac0002}. Through the help of these attack traces, an analyzer can detect the origin of the attack as well as determine the critical events and activities. Even if the adversary intentionally deletes some attack traces, the analyzer can perform \textit{forward causal analysis} to reconstruct those traces with the help of captured provenance records \cite{Hassan-Towards-18} \cite{Leeetal2013b}.

\par The captured attack traces contain information regarding the agents (e.g. users, groups) controlling the activities (e.g. processes) to interact with data objects (files, inodes, superblocks, socket buffers, IPC messages, IPC message queue, semaphores, and shared memory) \cite{Zhangetal2002} during system execution \cite{Batesetal2015}. If the provenance events, along with the agents and activities, are considered nodes, and the causal relationships among them are considered edges, the provenance traces can be depicted as a Directed Acyclic Graph (DAG) \cite{prov_dm}. The analyzer can leverage this graph to infer the causality and dependency of different system events that contribute to the compromise.

\par Because of continuous escalation of interest in system and data security, a lot of research efforts and advances have been made in provenance. These efforts can be roughly divided into three groups, namely provenance capture, management, and analysis. The provenance capture operations can be conducted at both system and user space levels. The capture mechanisms at the kernel space achieve their goals through either instrumenting the system calls \cite{Goeletal2005} \cite{Muniswamy-Reddy2009} or monitoring operations upon the kernel objects \cite{Pohlyetal2012} \cite{Batesetal2015}. The captured provenance records are at the granularity of process-level \cite{Pohlyetal2012} \cite{Batesetal2015} \cite{Batesetal2016} \cite{Pasquieretal2017} or subprocess-level \cite{Leeetal2013b} \cite{Maetal2017} \cite{Peietal2016} \cite{Maetal2016}. While the system level mechanisms can capture many details, special schemes must be used to map groups of provenance records back to semantics of system operations. 
Another group of capturing mechanisms focus on user level information, and allow applications to intercept user operations and events to generate provenance records \cite{Spillaneetal2009}. While these mechanisms usually provide more configuration power to users on granularity of captured information, integrity measures must be taken to make sure that the captured events are accurate and complete. 

\par Once the provenance records are captured, management and analysis operations can be conducted on the data. Restricted by the size of captured data, the management approaches often use offline methods to reduce the number of stored provenance records and redundant information in the records \cite{Leeetal2013b} \cite{Xuetal2016}. A limited number of research efforts concentrate on online (on-the-fly) provenance reduction \cite{Tangetal2018}. Similarly, data analysis and anomaly detection through provenance data can be conducted in either post-hoc methods \cite{Peietal2016} or real time schemes \cite{Pasquieretal2018a}. The detection capability and accuracy continue to improve with the fast development and adoption of machine learning and AI mechanisms in provenance.

\par In this paper, we provide a review of the approaches in these three categories and comparatively discuss their advantages and limitations. 
The remaining of the article is organized as follows. Section 2 presents the background knowledge and design choices of data provenance schemes. In Section 3, we study the approaches in each category in detail. In Section 4, we discuss data provenance in the emerging IoT domain, and in Section 5, we talk about the security issues of provenance system itself. Later, in Section 6, we introduce several research directions that deserve more efforts from the research community. And finally, we conclude the paper with Section 7.  


\section{Background}
\par To reconstruct an attack scenario and determine the critical attack nodes, the provenance records are expected to be accurate and complete enough to facilitate the reconstruction of system traces to infer sufficient information about events. Though a sophisticated attacker is capable to alter a provenance capture mechanism, a good provenance system should ensure the tamperproofness of this mechanism \cite{Batesetal2015} \cite{Pohlyetal2012}. In this section, we present the design choice of a good provenance system and conceptualize the components of provenance capture mechanism, provenance storage mechanism, and provenance analysis mechanism. 

\subsection{Design Choice}
\par The usability of provenance records depends on the design choice of the provenance system that ensures the trustworthiness and integrity of the captured records. In \cite{McDaniel2010} \cite{Batesetal2015}, the authors summarized the following four properties to ensure the usability of provenance records: (1) Reference Monitor Concept; (2) Traces Reproducibility; (3) Attested Disclosure; 
and (4) Network Authenticity. Below we will discuss the properties in detail.


\subsubsection{Reference Monitor Concept}
\par Reference monitor concept enforces the authorized access relationships between system subjects (e.g., users and processes that access the system resources) and objects (e.g. data, files, sockets, or subjects being accessed by the other subjects) based on an access control policy defined on \textit{reference validation mechanism}. It is used to explicitly control each subject\textquotesingle s access to any system resource that is shared with other subjects. It ensures that no subject practices over-privilege (e.g. READ, WRITE, EXECUTE) on any system object or resource \cite{Anderson1972} \cite{Jaeger2011}. Since the provenance system itself can become the target of malicious attacks, it is imperative to make sure that the execution of provenance will not incur security violations.

\par A reference monitor concept maintains the following three properties:

\begin{itemize}
    \item \textbf{Tamperproofness:} The reference validation mechanism must be tamperproof. The system should always behave correctly as expected.
    \item \textbf{Complete Mediation:} The reference validation mechanism must mediate every possible access initiated by the kernel.
    \item \textbf{Verifiability:} The reference validation mechanism must be small enough to be verified.
\end{itemize}

\par A good provenance system should be built around this reference monitor concept. LPM \cite{Batesetal2015} is an example of this. It is built on top of Linux Security Module (LSM) \cite{Wrightetal2003}, where a number of provenance hooks are placed carefully to mediate each kernel object access. LPM ensures the tamperproofness of the captured records through the deployment of SELinux MLS Policy \cite{Hicksetal2007}. The verification of the LPM is automatically maintained due to the deployment of LSM. As each provenance hook of LPM follows each LSM authorization hook, the correctness of LSM hook placement automatically verifies the LPM hooks. There are many techniques to verify the correctness of LSM hook placement. One such technique is described by Edwards et al. \cite{Edwardsetal2002}, where they used both static and dynamic analysis to verify the correctness.  



\subsubsection{Traces Reproducibility}
\par To undermine the derivation of attack behavior by the analyzer, adversaries may intentionally delete some important attack traces after successful attacks. Therefore, a good provenance system should ensure the reconstruction of missing attack traces to a certain extent so that these traces can be utilized in determining the attack behavior and understanding the level of damage.



\subsubsection{Attested Disclosure} 
\par Provenance recorder collects application-level semantic information from provenance-aware applications and incorporate it with kernel-level provenance records captured in kernel space. Hence, the low integrity of user space applications is a matter of concern. Therefore, a good provenance system should ensure the integrity of the applications prior to the disclosure of user-level provenance records. The applications should never be able to modify themselves when they are already loaded into memory. In this circumstance, Integrity Measurement Architecture (IMA) \cite{Saileretal2004} guarantees the integrity check of the applications in Linux kernel \cite{Batesetal2015}. 



\subsubsection{Network Authenticity} 
\par In a distributed environment, where provenance records are transmitted over the network, the provenance system must authenticate each possible outbound packet. One possible way to achieve this goal is to utilize a DSA signature to sign each outbound network packet prior to transmission, which should be verified immediately after receiving the packets at receiving end \cite{Batesetal2015}.


\subsection{A Conceptual System Overview}
\par
\begin{figure}[h]
    \centering
        \includegraphics[width = \textwidth]{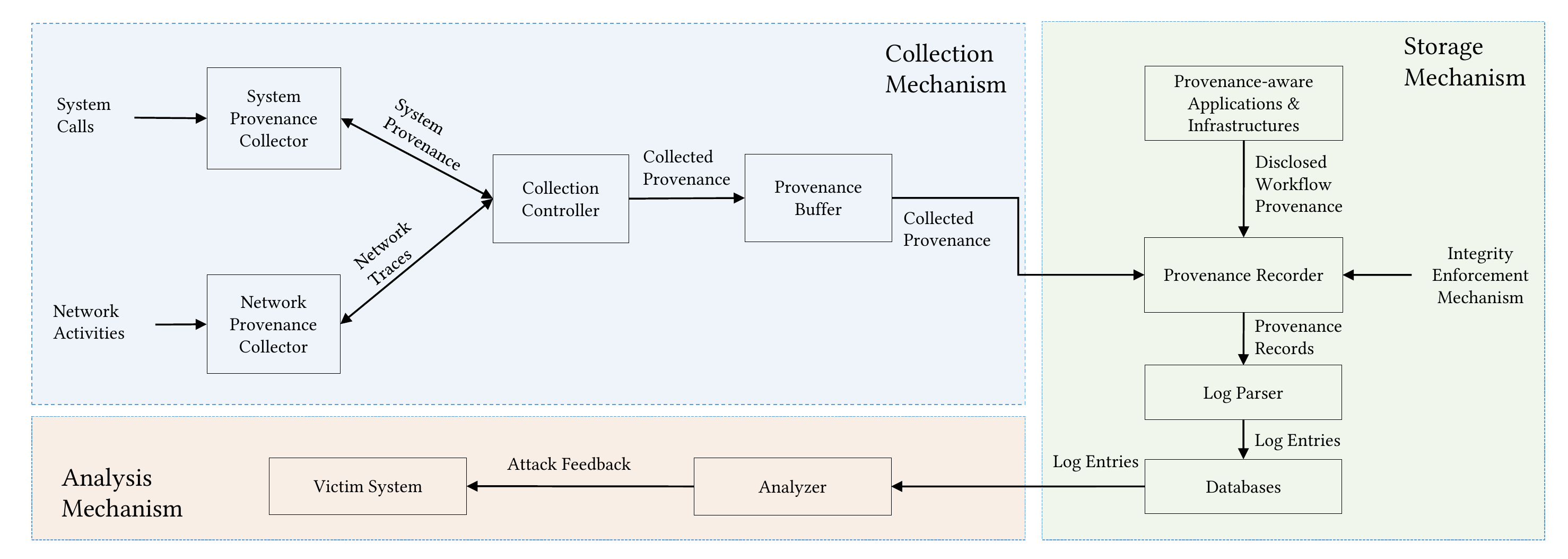}
    \caption{A conceptual view of a provenance system architecture}
    \label{FIG:SystemArchitecture}
\end{figure}

 As shown in Figure \ref{FIG:SystemArchitecture}, a conceptual provenance system consists of the following three components: (1) Provenance collection component; (2) Provenance storage component; and (3) Provenance analysis component. 
The figure visually explains how data flows from the point of capture to the storage and integrity assessment, and eventually to the analysis.
Although provenance collection conveys a larger meaning than just provenance capture, in this paper, we use the terms \textit{collection} and \textit{capture} interchangeably to denote the same thing for the readers' convenience throughout the rest of the paper. 

\subsubsection{Provenance Collection Component}
\par Provenance collection mechanism deals with the capture of provenance records in kernel space, the caching of records temporarily in a small buffer, and the transfer of the captured records into user space. Our observation is that the whole collection mechanism consists of the following three components:



\paragraph{i. Provenance Collector}
\par The existing literature adopted different kinds of approaches for kernel space provenance collection, including system call instrumentation \cite{Goeletal2005} \cite{Muniswamy-Reddy2009}, provenance and \textit{Netfilter} \cite{netfiter} hooks deployment \cite{Pohlyetal2012} \cite{Batesetal2015}, and \textit{Tracepoints} \cite{Tracepoints} deployment \cite{Maetal2016}. Different types of hooks and \textit{Tracepoints} are basically deployed to monitor the kernel objects and wait for any operations upon them to record the provenance data, while system calls are instrumented to directly attribute the processes, users, or connections.


\paragraph{ii. Collection Controller}
\par Collection controller is an entity that registers the collectors (e.g. provenance hooks, \textit{Tracepoints}) and regulates their operations. It performs file versioning to avoid cycles in provenance graphs, generates access control policy for the distributed system, and assigns random identifier for each outbound packet \cite{Batesetal2015}.


\paragraph{iii. Provenance Buffer}
\par A provenance buffer is a fixed-length storage space utilized for the exportation of captured provenance from kernel space to user space. Whenever the provenance collectors capture a provenance record, the collection controller immediately passes it to the provenance buffer. Then, the buffer exports it to the provenance recorder in user space for further operations. One such buffer is \textit{relayfs} \cite{Huttonetal2003}, which is used by LPM \cite{Batesetal2015} and CamFlow \cite{Pasquieretal2017}.


\subsubsection{Provenance Storage Component}
\par When provenance records are exported to the user space, the provenance storage mechanism converts them into suitable log entries prior to the storage operation. It performs integrity assessment of the provenance-aware applications when they intend to disclose workflow semantics. The whole mechanism consists of the following three components:

\paragraph{i. Provenance Recorder}
\par Provenance recorder facilitates the storage of provenance records to either local storage space or external server through the help of \textit{log parser}. It receives captured provenance records from the provenance buffer and disclosed workflow semantics from provenance-aware applications. In the case of disclosed provenance, it first ensures the integrity of the provenance-aware applications with the help of \textit{integrity enforcement mechanism} \cite{Saileretal2004}. The main purpose of this mechanism is to generate a cryptographic hash of each binary that is computed prior to each execution. This hash is used by the recorder to make a decision about the integrity of a provenance-aware application prior to accepting the disclosed provenance.


\paragraph{ii. Provenance-Aware Applications}
\par Provenance-aware applications report workflow semantics at a more abstract task structure than the provenance recorder. This reporting is conducted in the form of metadata, which is referred to as \textit{disclosed provenance} in the literature. \cite{Muniswamy-Reddy2009} \cite{Batesetal2015}. 


\paragraph{iii. Log Parser}
\par Provenance recorder forwards the provenance traces to the log parser to convert them into log entries so that the analyzer can easily perform queries over those entries. The log entries are further stored in databases in different formats, including Gzip, PostGreSQL, Neo4j, and SNAP \cite{Batesetal2015}.   




\subsubsection{Provenance Analysis component}
\par The provenance analysis mechanism takes log entries as input, performs suitable queries over the entries, and outputs necessary attack information. It explains how closely different system events (both benign and malicious events) are interrelated \cite{Peietal2016}, and what is the underlying behaviour of an attack.

\par The main component of this mechanism is the \textit{provenance analyzer}, which is generally invoked by the users when they experience unusual activities in the system. Its primary task is to provide attack feedback to the target system after determining the attack behavior of the adversary. It accesses stored log entries from the database and performs suitable queries over those entries. That's why its performance is bounded to the number of database entries and their formats.



\subsection{Provenance Representation}
\par World Wide Web Consortium (W3C) proposes a conceptual data model named PROV-DM \cite{prov_dm} to facilitate the representation of provenance records that describe the system entities, agents, and corresponding activities. PROV-DM allows the generation of a \textit{provenance graph} to delineate the data flow scenario throughout the whole system and illustrate the dependencies among system entities. Hence, the provenance graph is a directed acyclic graph where the entities, agents, and activities are the graph nodes and the relationships among them are the graph edges. In an attack scenario, this graph helps the analyzer discover the compromised system nodes and determine the attack path. To explain how the graph representation of provenance records helps analyze an attack, let us consider the following smart home scenario:   

\begin{figure}[h]
    \centering
        \includegraphics[width = \textwidth]{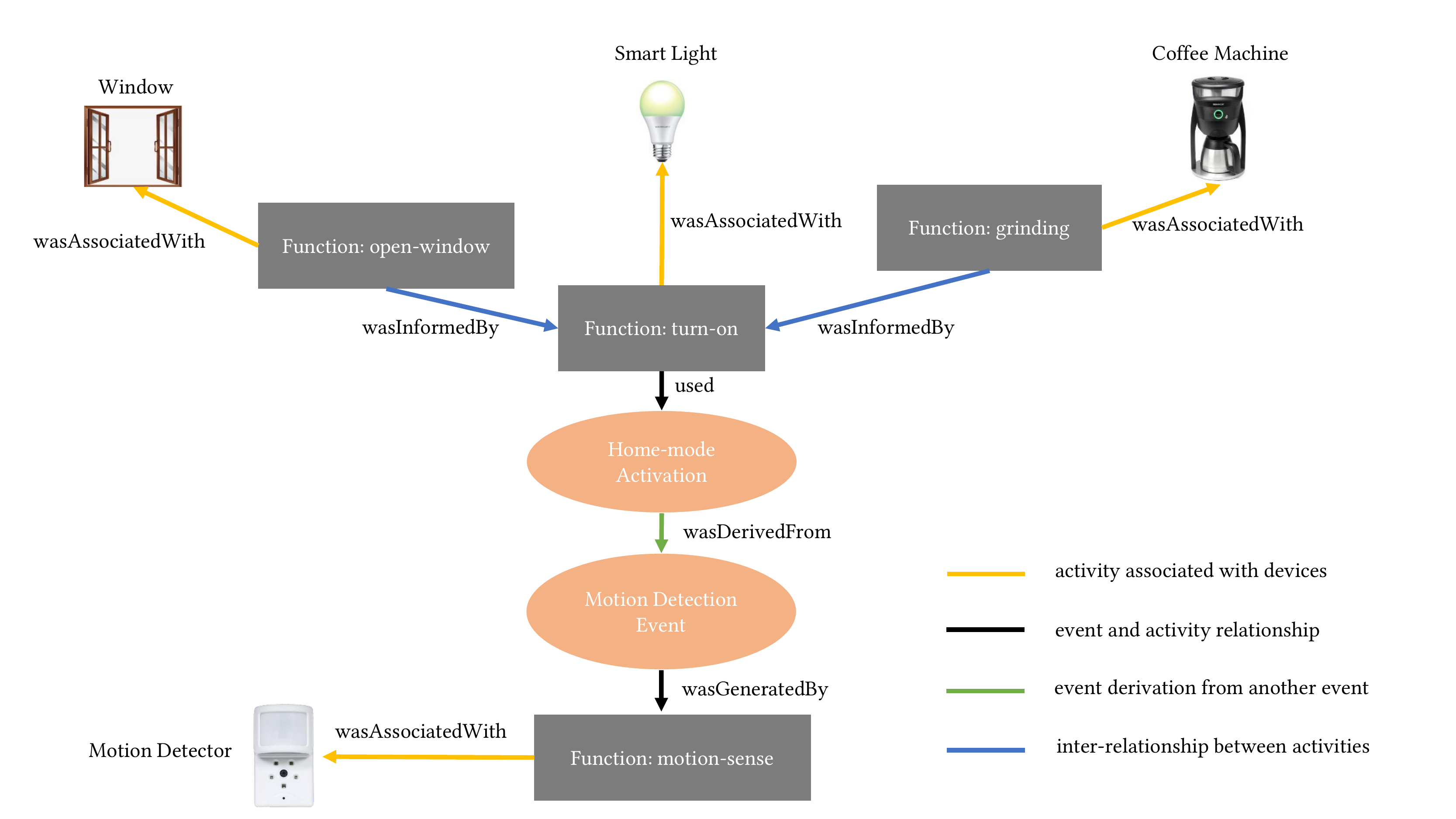}
    \caption{Provenance graph explaining Bob's trigger-action attack}
    \label{FIG:AttackScenario}
\end{figure}

\par Alice owns a smart home equipped with a smart lock, a motion detector, a smart light, a smart window, and a smart coffee machine. Alice controls these devices with mobile applications that function based on different trigger-action rules. Bob, on the contrary, is an attacker who is interested in adversely manipulating the actions of the devices. Alice always uses the smart lock mobile application to open the front door when she returns home from outside. When the door is unlocked, and she walks through the living room, the motion detector senses her motion and activates the \textit{home-mode}. This home-mode activation event triggers the \textit{turn-on} event associated with the smart light. When the light is turned on, the window automatically opens, and the coffee machine starts grinding coffee immediately. Now, Bob knows the exact locations of the devices and intentionally sets up his own devices just outside the home, which have greater sensing capability and computational power compared to Alice's devices. His ultimate goal is to open the window and start the coffee machine maliciously. Let us assume that, his devices somehow generate a fake motion detection event that activates the \textit{home-mode} and subsequently triggers the \textit{turn-on} event of smart light. If he becomes successful in triggering this \textit{turn-on} event, he can easily open the window and start the coffee machine to grind coffee.

\par Figure \ref{FIG:AttackScenario} depicts how data flows into different entities in the aforementioned scenario given that Bob becomes successful in manipulating the actions of the smart window and the coffee machine. Please note that the ovals represent the events that have occurred, and the rectangles represent the activities that are triggered during Bob's attack. The edges of the graph tell us how different events were produced or derived as well as how activities were generated and utilized. For example, through fooling the motion sensor (1), the attacker activates the "home-mode" (2), and turn on the light (3).

\par In addition to PROV-DM, W3C also proposes several provenance representation schemes, including PROV-XML \cite{Prov_XML}, PROV-JSON \cite{Prov_JSON}, and PROV-O \cite{Prov_Ontology}. 
\par In contrast PROV ontology, Cyber-investigation Analysis Standard Expression (CASE) \cite{CASEY201714} offers more flexibility to represent links and associations between objects using \textit{Relationship} objects. It also provides necessary functionalities to specify inputs, outputs, and the instruments used in \textit{Action}. In CASE, the result of an Action can also be another Action, which is not possible to cover in PROV ontology.




\section{Categorization and Comparative Analysis of Existing Approaches}
\par
\begin{figure}[h]
    \centering
        \includegraphics[width=\textwidth]{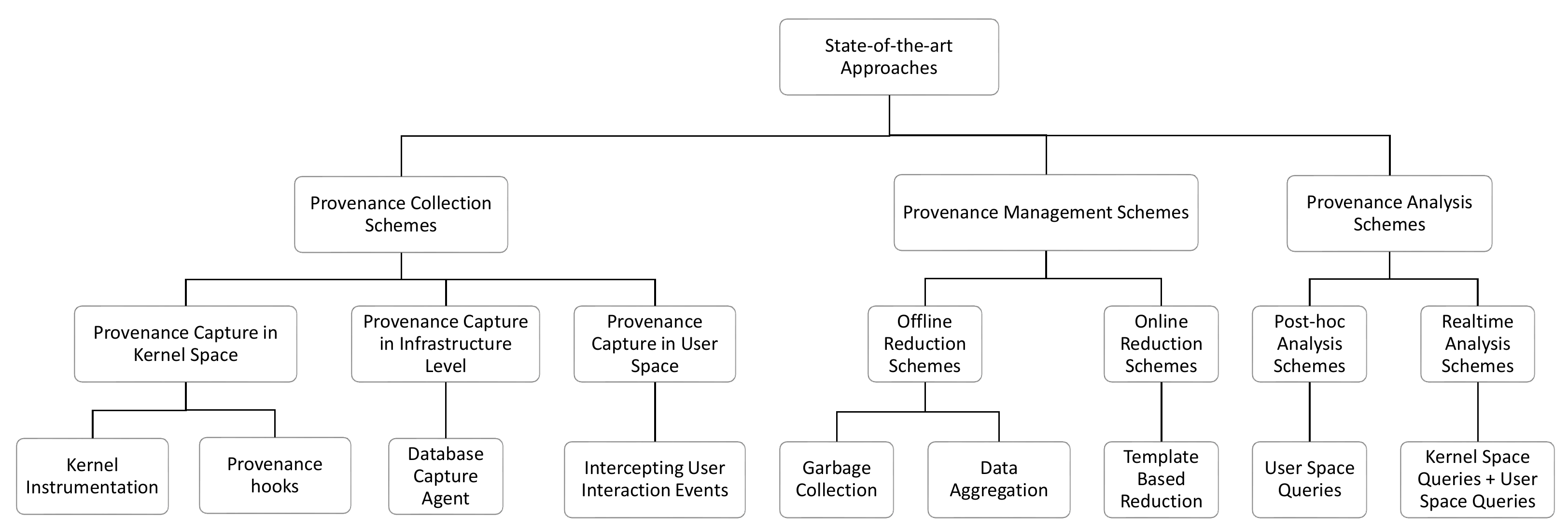}
    \caption{Categorization of the state-of-the-art approaches}
    \label{Fig:Categorization}
\end{figure}

 In this section, we comparatively study existing approaches to provenance collection, management, and analysis. The complete categorization of the state-of-the-art approaches is depicted in Figure ~\ref{Fig:Categorization}.

\subsection{Provenance Collection Schemes}
\par Existing literature includes a number of research efforts on provenance collection schemes. They mostly deal with event capturing with an emphasis on the optimization of reducing granularity of captured data. The detailed comparison among these mechanisms is illustrated in Table 1.

\subsubsection{Event Capturing Mechanism}
\par The state-of-the-art provenance collection schemes mostly capture either kernel-level provenance records in kernel space or semantic-aware provenance with application-level task structures in user space or at the infrastructure level \cite{Pasquieretal2018b} \cite{Braunetal2006}.

\par The earliest research efforts in the literature, including Forensix \cite{Goeletal2005}, PASSv1  \cite{Muniswamy-Reddy2006}, PASSv2 \cite{Muniswamy-Reddy2009}, and SPADE \cite{GehaniTariq2012}, directly instrument or intercept system calls to capture provenance records in kernel space. Since they only deal with the system calls, they are unable to capture all the non-persistent objects required to ensure \textit{whole-system provenance}. Here the missed non-persistent objects include IPC messages, socket buffers, shared memory, and the other objects that deal with the controlled data types defined by Zhang et al. \cite{Zhangetal2002}. 

\par To address this issue, Pohly et al. \cite{Pohlyetal2012} introduce Hi-Fi that captures high fidelity whole-system provenance starting from early kernel initialization through system shutdown. Hi-Fi leverages Linux Security Modules (LSM) \cite{Wrightetal2003} to mediate the access to system objects. It is built around \textit{reference monitor concept}, and it satisfies the design goals we presented in Section 2. 

\par In contrast to Hi-Fi, LPM \cite{Batesetal2015} creates a Trusted Computing Base (TCB) to collect whole-system provenance. It places 170 provenance hooks, one for each of the LSM authorization hook, to observe the system events and capture the provenance records. In addition, it uses several Netfilter hooks \cite{netfiter} to facilitate secure network transmission by implementing a cryptographic message commitment protocol. 

\begin{table}[h!]
\caption{Comparison among the Provenance Collection Schemes}
\label{table: Provenance Collection Schemes}
\centering
\begin{tabular}{@{} c c c c c @{}} 
 \hline
 Frameworks & \begin{tabular}[c]{@{}c@{}} Provenance \\ Collection Techniques\end{tabular} & \begin{tabular}[c]{@{}c@{}} Capturing \\ Non-persistent Objects\end{tabular} & \begin{tabular}[c]{@{}c@{}} Disclosed Provenance \\ Support\end{tabular} & Data Granularity \\ [0.5ex] 
 \hline
 
 Forensix \cite{Goeletal2005} & Kernel instrumentation & \begin{tabular}[c]{@{}c@{}} Does not \\ capture \end{tabular} & No & Instruction level \\ 
    
 SPADE \cite{GehaniTariq2012} & \begin{tabular}[c]{@{}c@{}} System-call \\ instrumentation \end{tabular} & -- & Yes & Instruction level \\
    
 PASSv1 \cite{Muniswamy-Reddy2006} & \begin{tabular}[c]{@{}c@{}} System-call \\ interception \end{tabular} & Pipes & No & File level \\
    
 PASSv2 \cite{Muniswamy-Reddy2009} & \begin{tabular}[c]{@{}c@{}} System-call \\ interception \end{tabular} & Pipes & Yes & File level \\
    
 Story Book \cite{Spillaneetal2009} & \begin{tabular}[c]{@{}c@{}c@{}c@{}} Intercepts \\ user interaction \\ with \\ application data \end{tabular} & -- & Yes & Query level* \\
    
 Hi-Fi \cite{Pohlyetal2012} & LSM Hooks & \begin{tabular}[c]{@{}c@{}} Whole-system \\ provenance \end{tabular} & No & Process level \\
    
 LPM \cite{Batesetal2015} & \begin{tabular}[c]{@{}c@{}} Trusted \\ Computing Base \end{tabular} & \begin{tabular}[c]{@{}c@{}} Whole-system \\ provenance \end{tabular} & Yes & Process level \\
    
 DAP \cite{Batesetal2016} & \begin{tabular}[c]{@{}c@{}} Database \\ Capture Agent \end{tabular} & \begin{tabular}[c]{@{}c@{}} Numbers, \\ string literals \end{tabular} & Yes & Process level \\
    
 CamFlow \cite{Pasquieretal2017} & \begin{tabular}[c]{@{}c@{}} LSM and \\ Netfilter hooks \end{tabular} & \begin{tabular}[c]{@{}c@{}} Whole-system \\ provenance \end{tabular} & Yes & Process level \\
    
 BEEP \cite{Leeetal2013a} & Reverse engineering & \begin{tabular}[c]{@{}c@{}} Memory, pipe, \\ socket \end{tabular} & No & \begin{tabular}[c]{@{}c@{}c@{}c@{}c@{}} Execution unit level \\ (based on \\ event handling loops) \end{tabular}  \\ \\
    
 Protracer \cite{Maetal2016} & Tracepoints & \begin{tabular}[c]{@{}c@{}} IPC, sockets, \\ pipes \end{tabular} & Yes & \begin{tabular}[c]{@{}c@{}c@{}c@{}c@{}} Execution unit level \\ (based on \\ event handling loops) \end{tabular} \\ \\
    
 MPI \cite{Maetal2017} &  \begin{tabular}[c]{@{}c@{}c@{}} Semantics aware \\ program annotation \\ and instrumentation \end{tabular} & \begin{tabular}[c]{@{}c@{}} IPC, sockets, \\ pipes \end{tabular} & -- & \begin{tabular}[c]{@{}c@{}c@{}c@{}c@{}} Execution unit level \\ (based on \\ high-level data structures) \end{tabular} \\ [1ex] 
 \hline
\end{tabular}
\end{table}


\par Similar to LPM, CamFlow \cite{Pasquieretal2017} also leverages LSM and Netfilter hooks to capture observed provenance records. It is proposed to support the integration of provenance across distributed systems and to minimize the overhead compared to Hi-Fi and LPM. It allows users to define the scope of provenance capture according to their own requirements. Users can decide whether CamFlow will capture whole-system provenance or selective provenance. If users choose selective-provenance, they can specify individual or combinations of the following criteria: (1) filters on nodes and edges; (2) specific programs and directories; (3) specific processes; and (4) specific network activities.

\par In addition to the techniques described above, there are several other capturing mechanisms, such as the deployment of \textit{Tracepoints} \cite{Tracepoints} and reverse engineering to capture low-level provenance data \cite{Leeetal2013a}. Protracer \cite{Maetal2016} places Tracepoints in both kernel and user code to provide hooks to call kernel functions. It performs event logging when an operation has permanent effect on the system entities (e.g. when a file is written, or a packet is sent through a socket). Otherwise, it performs unit-level taint propagation for intra-process information flow. Protracer alternates between event logging and taint-propagation to reduce the volume of captured records and to achieve fine granularity. BEEP \cite{Leeetal2013a}, on the other hand, reverse engineers the program loops from the application binaries. It reverse engineers the instructions that cause workflows between process units. 

\par Since kernel-level provenance data is difficult to interpret because of the semantic gap between kernel space and user space, there is always a demand of infrastructure-level or user space provenance capture mechanism in the application domain. Story Book \cite{Spillaneetal2009} facilitates the provenance capture in user space by intercepting the events of user interactions with application data and by sending these events to application specific extensions to interpret and generate provenance records of task level structures. As Story Book generates provenance data based on the user interactions with applications, it stores the generated provenance records separately from the application data. 

\par DAP \cite{Batesetal2016} is an add-on of LPM which is deployed at the infrastructure-level to capture semantically rich workflow provenance. The capture agent is a multi-threaded TCP proxy server that listens on the database engine's assigned ports. It extracts the database queries issued by the web applications and passes them through a Bison parser \cite{gnu_bison}. Then, it inspects the list of database objects that are accessed and creates a provenance event for each object.  

\subsubsection{Data Granularity}
\par The granularity of provenance data captured by instrumenting or intercepting system calls is, at best, instruction-level \cite{Goeletal2005} \cite{GehaniTariq2012} or file-level \cite{Muniswamy-Reddy2006} \cite{Muniswamy-Reddy2009}. On the contrary, the granularity of provenance data captured by the schemes based on provenance hooks is at process-level. Hi-Fi \cite{Pohlyetal2012}, LPM \cite{Batesetal2015}, DAP \cite{Batesetal2016}, and CamFlow \cite{Pasquieretal2017} are a few examples. However, these schemes frequently suffer from \textit{dependence explosion problem} that basically arises when a large volume of inputs and outputs are associated with a long running process, and each output of the process is causally dependent on all previous inputs. It emerges if a process has a non-trivial lifetime and inputs/outputs are repeatedly processed \cite{Leeetal2013a}.  

\par To address this issue and achieve finer granularity, BEEP \cite{Leeetal2013a} and Protracer \cite{Maetal2016} partition the execution of a process into multiple individual execution units depending on  kernel-level programming paradigms named \textit{event} \textit{handling} \textit{loops}. It results in event graphs with a lot of redundancy which are storage ineffective. Moreover, these schemes require prior training to detect memory dependencies across partitions, which is hard to achieve. MPI \cite{Maetal2017} is, therefore, introduced to address these issues. It allows a semantic program annotation and instrumentation technique to partition executions based on the application specific task structures. It first prompts an annotation miner that helps user annotate the program source code. Then, LLVM \cite{LLVM} pass takes the user annotations and analyzes the program to determine the data structures to instrument. Hence, the annotation miner is a data structure profiler and the LLVM pass is the analysis component.

\subsubsection{Layered Provenance Service}
\par Since the visibility of Operating System is limited to kernel space, it is mandatory for a provenance system to support a layered structure in order to achieve whole-system capture, where the provenance recorder receives workflow provenance from infrastructure and disclosed provenance from provenance-aware applications. In the existing literature, multiple provenance systems support this service, including PASSv2 \cite{Muniswamy-Reddy2009}, LPM \cite{Batesetal2015}, and CamFlow \cite{Pasquieretal2017}. 

\par However, there are some fundamental challenges related to the layering of provenance services, which are listed by Muniswamy-Reddy et al. \cite{Muniswamy-Reddy2009}. 
\begin{enumerate}
    \item Efficient communication among the provenance-aware components 
    \item Object identification and dependence extraction
    \item Maintenance of consistency semantics between data and provenance
    \item Cycle detection and removal
    \item Query support over provenance
    \item Security enforcement
\end{enumerate}

\par PASSv2 introduces Disclosed Provenance API (DPAPI) to allow the secure transfer of provenance among the components of the system and between layers. Whenever an object is accessed repeatedly, PASSv2 creates a new version of that object to avoid cycle. LPM also supports this type of file versioning for cycle avoidance. CamFlow, on the other hand, provides an API that allows the association of application provenance with system objects to ensure the avoidance of cycle as long as there is an available file descriptor to the application. In the case of integrity check of provenance-aware applications prior to the disclosure, LPM and CamFlow both use Linux IMA \cite{Saileretal2004}. 


\subsection{Provenance Management Schemes}
\par Provenance management schemes help reduce the size of collected provenance to save storage space and optimize query performance. In the literature, there are two types of management schemes - \textit{offline data reduction schemes} and \textit{online data reduction schemes}. Most of the introduced reduction schemes are offline schemes that work on log entries on storage media. The online schemes, on the contrary, deal with \textit{on-the-fly} data. For the reader's convenience, we will use the terms online and on-the-fly interchangeably throughout the rest of the paper. The detailed comparison among the offline and online schemes are shown in Table 2.  

\subsubsection{Offline Reduction Schemes}

\par LogGC \cite{Leeetal2013b} and Causality Preserved Reduction (CPR) \cite{Xuetal2016} are the two most common offline reduction approaches. LogGC is basically a garbage collection algorithm that discards log entries based on lifespan and influence over the dependency analysis. The main idea behind this algorithm is that many event entries are bound to the specific applications which are destroyed after the termination of these applications without further influencing any process or object. LogGC collects these entries as garbage since they are unreachable and not needed for future causal analysis. It leverages BEEP \cite{Leeetal2013a} to partition the execution of a process into multiple execution units to achieve fine granularity. LogGC is an application specific adaptation, and it requires human-in-the-loop to understand and change the specific applications. 

\par Causality Preserved Reduction (CPR) \cite{Xuetal2016}, on the other hand, leverages the dependencies among the system events to reduce the number of log entries. The key idea is that some events have identical contributions to the dependency analysis and can be aggregated \cite{Tanetal2018} without damaging inter-dependencies. It usually aggregates repeated  kernel level events (e.g. \textit{read} events) between two OS objects. Unlike LogGC, CPR is an enterprise adaptation, where there are hundreds of hosts containing thousands of diverse applications. 

\par In terms of reduction rate, LogGC reduces 92.89\% of the original audit logs for client systems and 97.35\% for server systems. On the other hand, CPR achieves a maximum of 77\% data reduction. CPR achieves less reduction compared to LogGC because CPR deals with a number of diverse applications that incur a diverse set of system events with unequal contributions to the dependency analysis. 

\subsubsection{Online Reduction Schemes}
\par NodeMerge \cite{Tangetal2018} is a template-based online data reduction scheme that achieves an additional 11.2\% reduction on the host level and 32.6\% reduction on the application level on top of Causality Preserved Reduction (CPR) \cite{Xuetal2016}. The key insight here is that some constant and intensive actions (e.g. loading libraries, accessing read-only resources, and retrieving configurations) are performed at every process initialization, and they can easily be grouped together without breaking the original data dependencies since they are \textit{read-only}, and they do not contain any useful system dependency information. NodeMerge generates templates based on the frequent access pattern of files and merges the incoming file events with the template files if there is a match. 

\par Protracer \cite{Maetal2016}, however, frequently switches between event logging and tainting based on the effect of operations. If the effect is permanent, it only logs. Otherwise, it performs unit-level taint propagation. For example, Protracer logs the events if there is a \textit{write} operation to a disk or a packet is sent through socket for either IPC or real network communication. In other cases, it performs taint propagation for any kind of intra-process operations. 

\par In terms of reduction rate, Protracer reduces even more than LogGC and CPR, since it avoids logging the redundant events, along with the dead-end events; while LogGC only removes the dead events and CPR aggregates the events based on the dependency.


\begin{table}[h!]
\caption{Comparison among the Provenance Reduction Schemes}
\label{Table: Reduction Scheme}
\centering
\begin{tabular}{@{} c c c c c @{}} 
 \hline
 Frameworks & Reduction Method & Type of Scheme & Scope & Reduction Rate \\ [0.5ex] 
 \hline
 LogGC \cite{Leeetal2013b} & \begin{tabular}[c]{@{}c@{}c@{}c@{}} Garbage Collection \\ based on \\ life-span and \\ dependency analysis \end{tabular} & Offline & Application-specific & 60.45\%-92.89\% \\ 
 
 CPR \cite{Xuetal2016} & \begin{tabular}[c]{@{}c@{}} Causality \\ preserved reduction \end{tabular} & Offline & \begin{tabular}[c]{@{}c@{}c@{}c@{}c@{}} Enterprise-specific\\ (Thousands of \\ diverse \\ applications) \end{tabular} & 56\%-77\% \\ 
    
 Protracer \cite{Maetal2016} & \begin{tabular}[c]{@{}c@{}} Logging and \\ tainting \end{tabular} & Online & Application-specific & >96\% \\ 
    
 NodeMerge \cite{Tangetal2018} & \begin{tabular}[c]{@{}c@{}} Merging \\ read-only files \end{tabular} & Online & Enterprise-specific & \begin{tabular}[c]{@{}c@{}} 11.2\% on the host level  \\ and 32.6\% on the \\ application level \\ on top of CPR \end{tabular}  \\ [1ex] 
 \hline
\end{tabular}
\end{table}



\subsection{Provenance Analysis Schemes}
\par Provenance analysis schemes leverage \textit{data relationship} to determine the root cause and impact of system compromise. In the existing literature, there are two types of provenance analysis schemes: post-hoc analysis schemes and runtime analysis schemes. 

\subsubsection{Post-hoc Analysis Scheme}
\par Post-hoc analysis schemes utilize stored provenance records to discover the relationships between system events and entities, and to determine the attack behavior. It is usually invoked after a system compromise, and it requires human-in-the-loop to perform the analysis. The recent advances in research lead to automated implementation of the tasks. One such example is Hercule \cite{Peietal2016}, which is an automated log-based intrusion analysis system that models the multi-stage intrusion analysis as a \textit{community discovery problem}. The ultimate goal of this approach is to discover the ``attack communities''. The key insight is that the attack logs are always heavily and densely connected with each other compared to the benign logs. Hercule performs causality analysis in modeling the relationship among multiple logs, and generates multi-dimensional weighted graphs. Then, it employs \textit{community detection algorithm} to extract the attack communities. The whole process described here takes place in user space after system compromise.

\subsubsection{Runtime Analysis Scheme}
\par Runtime analysis schemes, on the other hand, facilitate the timely response to a malicious incident in a system, which is crucial for a realtime security application. To thwart an ongoing attack, security measures are required to enforce in kernel space. Security enforcement in kernel-space ensures the tamperproofness of the collection method as well as the accuracy of the provenance records to perform analysis with. Runtime analysis schemes, such as CamQuery \cite{Pasquieretal2018a}, allow this type of enforcement by generating system level provenance graphs in kernel space, and by performing query analysis over the generated graphs. CamQuery specifically extends CamFlow \cite{Pasquieretal2017} to enable thread-level provenance capture. When provenance graphs are generated using the captured events, they are immediately fed into Loadable Kernel Module (LKM) to perform query analysis within the kernel space. After performing inline analysis, these graph elements are further transferred to the user space for post-hoc analysis or remote transmission.




\section{Provenance in IoT Devices}
\par The rapid emergence of IoT devices in both public and private spaces poses a real concern regarding the access and use of sensitive user data \cite{celik2019a} \cite{Fernandes2016}. Commodity IoT devices capture a wide range of user data in daily basis (e.g. door lock/unlock data, data related to energy consumption, user photo etc). An adversary may exploit these sensitive data to extract the daily habits of users, the pattern of their technology use, and the identifiable characteristics of them. Therefore, the handling  of these data should require the participation of end users in the process \cite{Pasquieretal2018b}. 

\par Sensitive user data including personal information, energy usage, locations frequently visited, personal habits, and physical conditions over a time can easily be captured by the commodity and health IoT devices, and user appliances. These types of data can be harnessed without users' permission to make important decisions regarding users' health insurance, credit card and other financial activities, employment, and utility services \cite{ftc2015} \cite{Elkhodr2019}.  Therefore, FTC staff report \cite{ftc2015} suggests to follow Fair Information Practice Principles (“FIPPs”) in handling captured users' data. This report focuses on four FIPPs in particular: 1) security; 2) data minimization; 3) notice; and 4) choice. Data provenance can be leveraged to ensure these FIPPs.

\par Data provenance can also be deployed to audit the flow of information in the IoT network and determine any security breach/risk. It can provide a detailed view of the network events and interactions across the whole network. It can also tell us whether any device, in particular, is having over-privilege (e.g. accessing location services always by a fitness tracker). Thus, the deployment of data provenance in IoT can ensure whether the devices in an IoT network are actually complying with the policies regarding information flow \cite{Pasquieretal2018b}. 

\par Due to the advent of trigger-action platforms (e.g. IFTTT \cite{IFTTT2020}), IoT devices create a chain of interactions maintaining functional dependencies between entities and actions \cite{Celik2019b}. Action of a certain device can be triggered because of the occurrence of another events(s) at another device. For instance, a window may open when a thermostat gives a measurement of 110\textdegree F. The chain of interactions include a number of this type of functional dependencies. Hence, data provenance comes handy in monitoring functional dependencies for a range of time period. It can also help determine the route to reach a certain device in the network. The use of provenance in IoT platforms, therefore, has the potential to enforce the privacy and security of the network \cite{Wangetal2018}.     

\par However, it is hard to collect whole system provenance in IoT because of the nature of this network. Since IoT devices are very light-weighted, and it is not imperative to impose computational overhead on the devices/sensors, fine-grained provenance is tough to achieve. Moreover, IoT network requires runtime provenance analysis, which is still in play.


\section{Security of Provenance Systems}
\par The earliest provenance systems, including SPADE \cite{GehaniTariq2012} and PASS \cite{Muniswamy-Reddy2006} were designed specifically for the benign environments. In contrast, the provenance systems built around \textit{reference monitor concept} were designed by assuming that the kernel can be compromised by an adversary. In the case of Hi-Fi \cite{Pohlyetal2012} and LogGC \cite{Leeetal2013b}, the adversary can tamper with the components of provenance collector if the kernel is compromised. However, LPM \cite{Batesetal2015} and CamFlow \cite{Pasquieretal2017} are resistant to this type of tampering because of the implementation of Trusted Computing Base (TCB).

\par When LogGC starts its operation, it assumes that all installed programs and files are clean. If, somehow, they are compromised prior to the installation of LogGC, it remains unable to detect the malicious activities. In that case, it produces deceptive provenance records. These types of records can be produced by any provenance system if it trusts all the already installed programs and existing files blindly. 




\par Although the partitioning schemes such as BEEP \cite{Leeetal2013a}, Protracer \cite{Maetal2016}, and MPI \cite{Maetal2017} solve the \textit{dependence explosion problem} by capturing provenance at execution unit-level granularity, their captured provenance suffers from the \textit{self-modification problem} due to the lack of integrity check, which may adversely affect the provenance recorder's intake of disclosed provenance from the applications and infrastructure.


\section{Future Directions}
\par Based on our discussion in the previous sections, it is quite evident that there is a demand for research efforts on online data reduction, runtime analysis, and several other aspects. In this section, we briefly introduce several future research directions that may emerge as intriguing in the future. 

\subsection{Noise Reduction in kernel level graphs}
\par In a system where execution partitioning schemes, such as BEEP \cite{Leeetal2013a} and Protracer \cite{Maetal2016}, are deployed, the system level provenance graphs generated in the kernel space contain a number of redundant nodes. This redundancy introduces noise in causal graphs that adversely impacts the performance of inline and realtime graph analysis \cite{Maetal2017}. Therefore, noise reduction in kernel level graphs deserves more research efforts.   


\subsection{Online Data Reduction}
\par In the case of APT (advanced persistent threat) attacks upon an organization, the causality analysis becomes prohibitively expensive and slow unless an on-the-fly reduction scheme removes a large portion of log entries prior to the storage operation. Typically, in a 3-4 month APT attack period upon a medium-sized enterprise, Peta Bytes of logs are generated \cite{Tangetal2018}. If all these logs are stored in the back end, the storage cost becomes undesirably high. Moreover, the generated causality graphs include a number of redundant dependencies that slow down the analysis procedure. In the existing literature, a limited number of research efforts have been conducted to reduce the captured volume on-the-fly, and therefore, there is ample room to delve into this domain.


\subsection{Leveraging Machine Learning into Provenance Analysis Techniques}
\par Machine learning algorithms can significantly impact the post-hoc analysis schemes. They can be deployed to determine the behavioral patterns of malicious activities. Similar to the deployment of \textit{community discovery algorithm} in Hercule \cite{Peietal2016}, other sophisticated machine learning algorithms can be used to discern the attack behavior. In runtime analysis schemes, machine learning models can be used to predict kernel level dependencies in order to analyze quickly the kernel space graphs.

\subsection{Leveraging Data Provenance to Audit Compliance with the Privacy Policy in Internet of Things}
\par Since commodity IoT devices capture sensitive user data, such as health information, user movement, user gesture, and user photo, the devices should always comply with the privacy policies for any kind of transactions. Users should know where and how their personal data are being used. They should always be notified about any changes in the usage policy. The data flows from device to device, or device to server should be transparent to the user. 

\par Pasquier et al. in \cite{Pasquieretal2018b} recommend the deployment of data provenance to audit compliance with the privacy policy in IoT. Data provenance may capture the inter-dependencies between IoT devices and applications. It may provide the holistic view of the system activities \cite{Wangetal2018} that can be facilitated to audit compliance with the privacy policy. Some approaches are needed to deal with accurate extraction of inter-dependencies, while others should concentrate on the deployment of this inter-dependency into policy-level.


\section{Conclusion}
\par Data provenance provides us with the comprehensive history of data and processes. We can leverage data provenance to reconstruct attack traces to learn how malicious activities propagate through a system. Traditionally, provenance is captured in the kernel space by instrumenting system calls or installing system hooks. It is arguably true that a well-crafted capture mechanism built around the \textit{reference monitor concept} can capture whole system provenance, including both bootstrap provenance and shutdown provenance. However, semantic program annotation and instrumentation techniques can be leveraged to achieve more enriched provenance with application specific task structures. The captured provenance records should necessarily pass through an online data reduction scheme to enable low-cost and efficient post-hoc analysis. Though an efficient post-hoc analysis is required to derive the attack behavior of an adversary, it is evident that we need inline and realtime provenance analysis techniques for the realtime security applications to counter ongoing attacks. In this article, we discuss different research approaches on provenance capture, management, and analysis. We believe that more research efforts on \textit{on-the-fly} data reduction and runtime analysis will be made in future. In conclusion, we recommend some probable research topics that may enrich the existing literature. 





\nocite{*} 
\bibliographystyle{ios1}           
\bibliography{bibliography}        

\begin{thebibliography}{57}
\ifx \bisbn   \undefined \def \bisbn  #1{ISBN #1}\fi
\ifx \binits  \undefined \def \binits#1{#1} \fi
\ifx \bauthor  \undefined \def \bauthor#1{#1} \fi
\ifx \bjtitle  \undefined \def \bjtitle#1{\textit{#1}}\fi
\ifx \batitle  \undefined \def \batitle#1{#1} \fi
\ifx \bctitle  \undefined \def \bctitle#1{#1} \fi
\ifx \bvolume  \undefined \def \bvolume#1{\textbf{#1}}\fi
\ifx \byear  \undefined \def \byear#1{#1} \fi
\ifx \bissue  \undefined \def \bissue#1{#1} \fi
\ifx \bfpage  \undefined \def \bfpage#1{#1} \fi
\ifx \blpage  \undefined \def \blpage #1{#1} \fi
\ifx \burl  \undefined \def \burl#1{#1} \fi
\ifx \doiurl  \undefined \def \doiurl#1{#1} \fi
\ifx \betal  \undefined \def \betal{et al.} \fi
\ifx \binstitute  \undefined \def \binstitute#1{#1} \fi
\ifx \beditor  \undefined \def \beditor#1{#1} \fi
\ifx \bpublisher  \undefined \def \bpublisher#1{#1} \fi
\ifx \bbtitle  \undefined \def \bbtitle#1{\textit{#1}} \fi
\ifx \bedition  \undefined \def \bedition#1{#1} \fi
\ifx \bseriesno  \undefined \def \bseriesno#1{#1} \fi
\ifx \blocation  \undefined \def \blocation#1{#1} \fi
\ifx \bsertitle  \undefined \def \bsertitle#1{#1} \fi
\ifx \bsnm \undefined \def \bsnm#1{#1} \fi
\ifx \bsuffix \undefined \def \bsuffix#1{#1} \fi
\ifx \bparticle \undefined \def \bparticle#1{#1} \fi
\ifx \barticle \undefined \def \barticle#1{#1} \fi
\ifx \botherref \undefined \def \botherref #1{#1} \fi
\ifx \url \undefined \def \url#1{#1} \fi
\ifx \bchapter \undefined \def \bchapter#1{#1} \fi
\ifx \bbook \undefined \def \bbook#1{#1} \fi
\ifx \bcomment \undefined \def \bcomment#1{#1} \fi
\ifx \oauthor \undefined \def \oauthor#1{#1} \fi
\ifx \citeauthoryear \undefined \def \citeauthoryear#1{#1} \fi
\ifx \texttildelow  \undefined \def \texttildelow{\symbol{126}} \fi
\def \endbibitem {}
\ifx \bconflocation  \undefined \def \bconflocation#1{#1} \fi

\bibitem{Bunemanetal2001}
\begin{bchapter}
\bauthor{\binits{P.}~\bsnm{Buneman}},
\bauthor{\binits{S.}~\bsnm{Khanna}} and
\bauthor{\binits{T.}~\bsnm{Wang-Chiew}},
\bctitle{Why and Where: A Characterization of Data Provenance},
in: \bbtitle{Database Theory - ICDT 2001},
\beditor{\binits{J.}~\bsnm{Van~den Bussche}} and
\beditor{\binits{V.}~\bsnm{Vianu}}, eds,
\bpublisher{Springer Berlin Heidelberg},
\blocation{Berlin, Heidelberg},
\byear{2001},
pp.~\bfpage{316}--\blpage{330}.
ISBN \bisbn{978-3-540-44503-6}.
\end{bchapter}
\endbibitem

\bibitem{Simmhanetal2005}
\begin{barticle}
\bauthor{\binits{Y.L.}~\bsnm{Simmhan}},
\bauthor{\binits{B.}~\bsnm{Plale}} and
\bauthor{\binits{D.}~\bsnm{Gannon}},
\batitle{A Survey of Data Provenance in e-Science},
\bjtitle{SIGMOD Rec.}
\bvolume{34}(\bissue{3})
(\byear{2005}),
\bfpage{31}--\blpage{36}.
doi:\doiurl{10.1145/1084805.1084812}.
\end{barticle}
\endbibitem

\bibitem{Zafar-Trust-17}
\begin{barticle}
\bauthor{\binits{F.}~\bsnm{Zafar}},
\bauthor{\binits{A.}~\bsnm{Khan}},
\bauthor{\binits{S.}~\bsnm{Suhail}},
\bauthor{\binits{I.}~\bsnm{Ahmed}},
\bauthor{\binits{K.}~\bsnm{Hameed}},
\bauthor{\binits{H.M.}~\bsnm{Khan}},
\bauthor{\binits{F.}~\bsnm{Jabeen}} and
\bauthor{\binits{A.}~\bsnm{Anjum}},
\batitle{Trustworthy data: A survey, taxonomy and future trends of secure
  provenance schemes},
\bjtitle{Journal of Network and Computer Applications}
\bvolume{94}
(\byear{2017}),
\bfpage{50}--\blpage{68}.
\end{barticle}
\endbibitem

\bibitem{osac0002}
\begin{botherref}
A Framework for Harmonizing Forensic Science Practices and Digital/Multimedia
  Evidence,
The Organization of Scientific Area Committees for Forensic Science (OSAC),
Accessed: 2020-11-17.
\end{botherref}
\endbibitem

\bibitem{Hassan-Towards-18}
\begin{bchapter}
\bauthor{\binits{W.U.}~\bsnm{Hassan}},
\bauthor{\binits{M.}~\bsnm{Lemay}},
\bauthor{\binits{N.}~\bsnm{Aguse}},
\bauthor{\binits{A.}~\bsnm{Bates}} and
\bauthor{\binits{T.}~\bsnm{Moyer}},
\bctitle{Towards Scalable Cluster Auditing through Grammatical Inference over
  Provenance Graphs},
in: \bbtitle{Network and Distributed System Security Symposium (NDSS)},
\byear{2018}.
\end{bchapter}
\endbibitem

\bibitem{Leeetal2013b}
\begin{bchapter}
\bauthor{\binits{K.H.}~\bsnm{Lee}},
\bauthor{\binits{X.}~\bsnm{Zhang}} and
\bauthor{\binits{D.}~\bsnm{Xu}},
\bctitle{LogGC: garbage collecting audit log},
in: \bbtitle{Proceedings of the 2013 ACM SIGSAC conference on Computer and
  communications Security},
\bsertitle{CCS '13},
\bpublisher{ACM},
\blocation{New York, NY, USA},
\byear{2013},
pp.~\bfpage{1005}--\blpage{1016}.
ISBN \bisbn{978-1-4503-2477-9}.
doi:\doiurl{10.1145/2508859.2516731}.
\end{bchapter}
\endbibitem

\bibitem{Zhangetal2002}
\begin{bchapter}
\bauthor{\binits{X.}~\bsnm{Zhang}},
\bauthor{\binits{A.}~\bsnm{Edwards}} and
\bauthor{\binits{T.}~\bsnm{Jaeger}},
\bctitle{Using CQUAL for Static Analysis of Authorization Hook Placement},
in: \bbtitle{Proceedings of the 11th USENIX Security Symposium},
\bpublisher{USENIX Association},
\blocation{Berkeley, CA, USA},
\byear{2002},
pp.~\bfpage{33}--\blpage{48}.
ISBN \bisbn{1-931971-00-5}.
\url{http://dl.acm.org/citation.cfm?id=647253.720279}.
\end{bchapter}
\endbibitem

\bibitem{Batesetal2015}
\begin{bchapter}
\bauthor{\binits{A.}~\bsnm{Bates}},
\bauthor{\binits{D.J.}~\bsnm{Tian}},
\bauthor{\binits{K.R.B.}~\bsnm{Butler}} and
\bauthor{\binits{T.}~\bsnm{Moyer}},
\bctitle{Trustworthy Whole-System Provenance for the Linux Kernel},
in: \bbtitle{24th {USENIX} Security Symposium ({USENIX} Security 15)},
\bpublisher{{USENIX} Association},
\blocation{Washington, D.C.},
\byear{2015},
pp.~\bfpage{319}--\blpage{334}.
ISBN \bisbn{978-1-931971-232}.
\url{https://www.usenix.org/conference/usenixsecurity15/technical-sessions/presentation/bates}.
\end{bchapter}
\endbibitem

\bibitem{prov_dm}
\begin{botherref}
\oauthor{\binits{K.}~\bsnm{Belhajjame}},
\oauthor{\binits{R.}~\bsnm{B'Far}},
\oauthor{\binits{J.}~\bsnm{Cheney}},
\oauthor{\binits{S.}~\bsnm{Coppens}},
\oauthor{\binits{S.}~\bsnm{Cresswell}},
\oauthor{\binits{Y.}~\bsnm{Gil}},
\oauthor{\binits{P.}~\bsnm{Groth}},
\oauthor{\binits{G.}~\bsnm{Klyne}},
\oauthor{\binits{T.}~\bsnm{Lebo}},
\oauthor{\binits{J.}~\bsnm{McCusker}},
\oauthor{\binits{S.}~\bsnm{Miles}},
\oauthor{\binits{J.}~\bsnm{Myers}},
\oauthor{\binits{S.}~\bsnm{Sahoo}} and
\oauthor{\binits{C.}~\bsnm{Tilmes}},
Prov-DM: The Provenance Data Model,
2013,
Accessed: 2018-04-30.
\end{botherref}
\endbibitem

\bibitem{Goeletal2005}
\begin{bchapter}
\bauthor{\binits{A.}~\bsnm{Goel}},
\bauthor{\binits{W.-c.}~\bsnm{Feng}},
\bauthor{\binits{D.}~\bsnm{Maier}},
\bauthor{\binits{W.-c.}~\bsnm{Feng}} and
\bauthor{\binits{J.}~\bsnm{Walpole}},
\bctitle{Forensix: a robust, high-performance reconstruction system},
in: \bbtitle{25th IEEE International Conference on Distributed Computing
  Systems Workshops},
\byear{2005},
pp.~\bfpage{155}--\blpage{162}.
\bcomment{ISSN 1545-0678}.
doi:\doiurl{10.1109/ICDCSW.2005.62}.
\end{bchapter}
\endbibitem

\bibitem{Muniswamy-Reddy2009}
\begin{bchapter}
\bauthor{\binits{K.-K.}~\bsnm{Muniswamy-Reddy}},
\bauthor{\binits{U.}~\bsnm{Braun}},
\bauthor{\binits{D.A.}~\bsnm{Holland}},
\bauthor{\binits{P.}~\bsnm{Macko}},
\bauthor{\binits{D.}~\bsnm{Maclean}},
\bauthor{\binits{D.}~\bsnm{Margo}},
\bauthor{\binits{M.}~\bsnm{Seltzer}} and
\bauthor{\binits{R.}~\bsnm{Smogor}},
\bctitle{Layering in Provenance Systems},
in: \bbtitle{Proceedings of the 2009 Conference on USENIX Annual Technical
  Conference},
\bsertitle{USENIX'09},
\bpublisher{USENIX Association},
\blocation{Berkeley, CA, USA},
\byear{2009},
pp.~\bfpage{10}--\blpage{10}.
\url{http://dl.acm.org/citation.cfm?id=1855807.1855817}.
\end{bchapter}
\endbibitem

\bibitem{Pohlyetal2012}
\begin{bchapter}
\bauthor{\binits{D.J.}~\bsnm{Pohly}},
\bauthor{\binits{S.}~\bsnm{McLaughlin}},
\bauthor{\binits{P.}~\bsnm{McDaniel}} and
\bauthor{\binits{K.}~\bsnm{Butler}},
\bctitle{Hi-Fi: Collecting High-fidelity Whole-system Provenance},
in: \bbtitle{Proceedings of the 28th Annual Computer Security Applications
  Conference},
\bsertitle{ACSAC '12},
\bpublisher{ACM},
\blocation{New York, NY, USA},
\byear{2012},
pp.~\bfpage{259}--\blpage{268}.
ISBN \bisbn{978-1-4503-1312-4}.
doi:\doiurl{10.1145/2420950.2420989}.
\end{bchapter}
\endbibitem

\bibitem{Batesetal2016}
\begin{botherref}
\oauthor{\binits{A.M.}~\bsnm{Bates}},
\oauthor{\binits{K.R.B.}~\bsnm{Butler}},
\oauthor{\binits{A.}~\bsnm{Dobra}},
\oauthor{\binits{B.}~\bsnm{Reaves}},
\oauthor{\binits{P.T.}~\bsnm{Cable}},
\oauthor{\binits{T.}~\bsnm{Moyer}} and
\oauthor{\binits{N.}~\bsnm{Schear}},
Retrofitting Applications with Provenance-Based Security Monitoring,
\textit{CoRR}
\textbf{abs/1609.00266}
(2016).
\url{http://arxiv.org/abs/1609.00266}.
\end{botherref}
\endbibitem

\bibitem{Pasquieretal2017}
\begin{bchapter}
\bauthor{\binits{T.}~\bsnm{Pasquier}},
\bauthor{\binits{X.}~\bsnm{Han}},
\bauthor{\binits{M.}~\bsnm{Goldstein}},
\bauthor{\binits{T.}~\bsnm{Moyer}},
\bauthor{\binits{D.}~\bsnm{Eyers}},
\bauthor{\binits{M.}~\bsnm{Seltzer}} and
\bauthor{\binits{J.}~\bsnm{Bacon}},
\bctitle{Practical Whole-system Provenance Capture},
in: \bbtitle{Proceedings of the 2017 Symposium on Cloud Computing},
\bsertitle{SoCC '17},
\bpublisher{ACM},
\blocation{New York, NY, USA},
\byear{2017},
pp.~\bfpage{405}--\blpage{418}.
ISBN \bisbn{978-1-4503-5028-0}.
doi:\doiurl{10.1145/3127479.3129249}.
\end{bchapter}
\endbibitem

\bibitem{Maetal2017}
\begin{bchapter}
\bauthor{\binits{S.}~\bsnm{Ma}},
\bauthor{\binits{J.}~\bsnm{Zhai}},
\bauthor{\binits{F.}~\bsnm{Wang}},
\bauthor{\binits{K.H.}~\bsnm{Lee}},
\bauthor{\binits{X.}~\bsnm{Zhang}} and
\bauthor{\binits{D.}~\bsnm{Xu}},
\bctitle{{MPI}: Multiple Perspective Attack Investigation with Semantic Aware
  Execution Partitioning},
in: \bbtitle{26th {USENIX} Security Symposium ({USENIX} Security 17)},
\bpublisher{{USENIX} Association},
\blocation{Vancouver, BC},
\byear{2017},
pp.~\bfpage{1111}--\blpage{1128}.
ISBN \bisbn{978-1-931971-40-9}.
\url{https://www.usenix.org/conference/usenixsecurity17/technical-sessions/presentation/ma}.
\end{bchapter}
\endbibitem

\bibitem{Peietal2016}
\begin{bchapter}
\bauthor{\binits{K.}~\bsnm{Pei}},
\bauthor{\binits{Z.}~\bsnm{Gu}},
\bauthor{\binits{B.}~\bsnm{Saltaformaggio}},
\bauthor{\binits{S.}~\bsnm{Ma}},
\bauthor{\binits{F.}~\bsnm{Wang}},
\bauthor{\binits{Z.}~\bsnm{Zhang}},
\bauthor{\binits{L.}~\bsnm{Si}},
\bauthor{\binits{X.}~\bsnm{Zhang}} and
\bauthor{\binits{D.}~\bsnm{Xu}},
\bctitle{HERCULE: Attack Story Reconstruction via Community Discovery on
  Correlated Log Graph},
in: \bbtitle{Proceedings of the 32Nd Annual Conference on Computer Security
  Applications},
\bsertitle{ACSAC '16},
\bpublisher{ACM},
\blocation{New York, NY, USA},
\byear{2016},
pp.~\bfpage{583}--\blpage{595}.
ISBN \bisbn{978-1-4503-4771-6}.
doi:\doiurl{10.1145/2991079.2991122}.
\end{bchapter}
\endbibitem

\bibitem{Maetal2016}
\begin{bchapter}
\bauthor{\binits{S.}~\bsnm{Ma}},
\bauthor{\binits{X.}~\bsnm{Zhang}} and
\bauthor{\binits{D.}~\bsnm{Xu}},
\bctitle{ProTracer: Towards Practical Provenance Tracing by Alternating Between
  Logging and Tainting},
in: \bbtitle{Network and Distributed System Security Symposium},
\byear{2016}.
doi:\doiurl{10.14722/ndss.2016.23350}.
\end{bchapter}
\endbibitem

\bibitem{Spillaneetal2009}
\begin{bchapter}
\bauthor{\binits{R.}~\bsnm{Spillane}},
\bauthor{\binits{R.}~\bsnm{Sears}},
\bauthor{\binits{C.}~\bsnm{Yalamanchili}},
\bauthor{\binits{S.}~\bsnm{Gaikwad}},
\bauthor{\binits{M.}~\bsnm{Chinni}} and
\bauthor{\binits{E.}~\bsnm{Zadok}},
\bctitle{Story Book: An Efficient Extensible Provenance Framework},
in: \bbtitle{First Workshop on on Theory and Practice of Provenance},
\bsertitle{TAPP'09},
\bpublisher{USENIX Association},
\blocation{Berkeley, CA, USA},
\byear{2009},
pp.~\bfpage{11:1}--\blpage{11:10}.
\url{http://dl.acm.org/citation.cfm?id=1525932.1525943}.
\end{bchapter}
\endbibitem

\bibitem{Xuetal2016}
\begin{bchapter}
\bauthor{\binits{Z.}~\bsnm{Xu}},
\bauthor{\binits{Z.}~\bsnm{Wu}},
\bauthor{\binits{Z.}~\bsnm{Li}},
\bauthor{\binits{K.}~\bsnm{Jee}},
\bauthor{\binits{J.}~\bsnm{Rhee}},
\bauthor{\binits{X.}~\bsnm{Xiao}},
\bauthor{\binits{F.}~\bsnm{Xu}},
\bauthor{\binits{H.}~\bsnm{Wang}} and
\bauthor{\binits{G.}~\bsnm{Jiang}},
\bctitle{High Fidelity Data Reduction for Big Data Security Dependency
  Analyses},
in: \bbtitle{Proceedings of the 2016 ACM SIGSAC Conference on Computer and
  Communications Security},
\bsertitle{CCS '16},
\bpublisher{ACM},
\blocation{New York, NY, USA},
\byear{2016},
pp.~\bfpage{504}--\blpage{516}.
ISBN \bisbn{978-1-4503-4139-4}.
doi:\doiurl{10.1145/2976749.2978378}.
\end{bchapter}
\endbibitem

\bibitem{Tangetal2018}
\begin{bchapter}
\bauthor{\binits{Y.}~\bsnm{Tang}},
\bauthor{\binits{D.}~\bsnm{Li}},
\bauthor{\binits{Z.}~\bsnm{Li}},
\bauthor{\binits{M.}~\bsnm{Zhang}},
\bauthor{\binits{K.}~\bsnm{Jee}},
\bauthor{\binits{X.}~\bsnm{Xiao}},
\bauthor{\binits{Z.}~\bsnm{Wu}},
\bauthor{\binits{J.}~\bsnm{Rhee}},
\bauthor{\binits{F.}~\bsnm{Xu}} and
\bauthor{\binits{Q.}~\bsnm{Li}},
\bctitle{NodeMerge: Template Based Efficient Data Reduction For Big-Data
  Causality Analysis},
in: \bbtitle{Proceedings of the 2018 ACM SIGSAC Conference on Computer and
  Communications Security},
\bsertitle{CCS '18},
\bpublisher{ACM},
\blocation{New York, NY, USA},
\byear{2018},
pp.~\bfpage{1324}--\blpage{1337}.
ISBN \bisbn{978-1-4503-5693-0}.
doi:\doiurl{10.1145/3243734.3243763}.
\end{bchapter}
\endbibitem

\bibitem{Pasquieretal2018a}
\begin{botherref}
\oauthor{\binits{T.}~\bsnm{Pasquier}},
\oauthor{\binits{X.}~\bsnm{Han}},
\oauthor{\binits{T.}~\bsnm{Moyer}},
\oauthor{\binits{A.}~\bsnm{Bates}},
\oauthor{\binits{O.}~\bsnm{Hermant}},
\oauthor{\binits{D.}~\bsnm{Eyers}},
\oauthor{\binits{J.}~\bsnm{Bacon}} and
\oauthor{\binits{M.}~\bsnm{Seltzer}},
{Runtime Analysis of Whole-System Provenance},
\textit{Computer and Communications Security (CCS)}
(2018).
\url{http://arxiv.org/abs/1808.06049}.
\end{botherref}
\endbibitem

\bibitem{McDaniel2010}
\begin{bchapter}
\bauthor{\binits{P.}~\bsnm{McDaniel}},
\bauthor{\binits{K.}~\bsnm{Butler}},
\bauthor{\binits{S.}~\bsnm{McLaughlin}},
\bauthor{\binits{R.}~\bsnm{Sion}},
\bauthor{\binits{E.}~\bsnm{Zadok}} and
\bauthor{\binits{M.}~\bsnm{Winslett}},
\bctitle{Towards a Secure and Efficient System for End-to-end Provenance},
in: \bbtitle{Proceedings of the 2Nd Conference on Theory and Practice of
  Provenance},
\bsertitle{TAPP'10},
\bpublisher{USENIX Association},
\blocation{Berkeley, CA, USA},
\byear{2010},
pp.~\bfpage{2}--\blpage{2}.
\url{http://dl.acm.org/citation.cfm?id=1855795.1855797}.
\end{bchapter}
\endbibitem

\bibitem{Anderson1972}
\begin{botherref}
\oauthor{\binits{J.P.}~\bsnm{Anderson}},
Computer Security Technology Planning Study,
Technical Report, ESD-TR-73-51,
U.S. Air Force Electronic Systems Division,
1972.
\url{https://csrc.nist.gov/csrc/media/publications/conference-paper/1998/10/08/proceedings-of-the-21st-nissc-1998/documents/early-cs-papers/ande72a.pdf}.
\end{botherref}
\endbibitem

\bibitem{Jaeger2011}
\begin{bchapter}
\bauthor{\binits{T.}~\bsnm{Jaeger}},
\bbtitle{Reference Monitor},
in: \bbtitle{Encyclopedia of Cryptography and Security},
\beditor{\binits{H.C.A.}~\bsnm{van Tilborg}} and
\beditor{\binits{S.}~\bsnm{Jajodia}}, eds,
\bpublisher{Springer US},
\blocation{Boston, MA},
\byear{2011},
pp.~\bfpage{1038}--\blpage{1040}.
ISBN \bisbn{978-1-4419-5906-5}.
\url{https://doi.org/10.1007/978-1-4419-5906-5_646}.
\end{bchapter}
\endbibitem

\bibitem{Wrightetal2003}
\begin{bchapter}
\bauthor{\binits{C.}~\bsnm{Wright}},
\bauthor{\binits{C.}~\bsnm{Cowan}},
\bauthor{\binits{J.}~\bsnm{Morris}},
\bauthor{\binits{S.}~\bsnm{Smalley}} and
\bauthor{\binits{G.}~\bsnm{Kroah-Hartman}},
\bctitle{Linux security modules: general security support for the linux
  kernel},
in: \bbtitle{Foundations of Intrusion Tolerant Systems, 2003 [Organically
  Assured and Survivable Information Systems]},
\byear{2003},
pp.~\bfpage{213}--\blpage{226}.
doi:\doiurl{10.1109/FITS.2003.1264934}.
\end{bchapter}
\endbibitem

\bibitem{Hicksetal2007}
\begin{bchapter}
\bauthor{\binits{B.}~\bsnm{Hicks}},
\bauthor{\binits{S.}~\bsnm{Rueda}},
\bauthor{\binits{L.}~\bsnm{St.Clair}},
\bauthor{\binits{T.}~\bsnm{Jaeger}} and
\bauthor{\binits{P.}~\bsnm{McDaniel}},
\bctitle{A Logical Specification and Analysis for SELinux MLS Policy},
in: \bbtitle{Proceedings of the 12th ACM Symposium on Access Control Models and
  Technologies},
\bsertitle{SACMAT '07},
\bpublisher{ACM},
\blocation{New York, NY, USA},
\byear{2007},
pp.~\bfpage{91}--\blpage{100}.
ISBN \bisbn{978-1-59593-745-2}.
doi:\doiurl{10.1145/1266840.1266854}.
\end{bchapter}
\endbibitem

\bibitem{Edwardsetal2002}
\begin{bchapter}
\bauthor{\binits{A.}~\bsnm{Edwards}},
\bauthor{\binits{T.}~\bsnm{Jaeger}} and
\bauthor{\binits{X.}~\bsnm{Zhang}},
\bctitle{Runtime Verification of Authorization Hook Placement for the Linux
  Security Modules Framework},
in: \bbtitle{Proceedings of the 9th ACM Conference on Computer and
  Communications Security},
\bsertitle{CCS '02},
\bpublisher{ACM},
\blocation{New York, NY, USA},
\byear{2002},
pp.~\bfpage{225}--\blpage{234}.
ISBN \bisbn{1-58113-612-9}.
doi:\doiurl{10.1145/586110.586141}.
\end{bchapter}
\endbibitem

\bibitem{Saileretal2004}
\begin{bchapter}
\bauthor{\binits{R.}~\bsnm{Sailer}},
\bauthor{\binits{X.}~\bsnm{Zhang}},
\bauthor{\binits{T.}~\bsnm{Jaeger}} and
\bauthor{\binits{L.}~\bsnm{van Doorn}},
\bctitle{Design and Implementation of a TCG-based Integrity Measurement
  Architecture},
in: \bbtitle{Proceedings of the 13th Conference on USENIX Security Symposium -
  Volume 13},
\bsertitle{SSYM'04},
\bpublisher{USENIX Association},
\blocation{Berkeley, CA, USA},
\byear{2004},
pp.~\bfpage{16}--\blpage{16}.
\url{http://dl.acm.org/citation.cfm?id=1251375.1251391}.
\end{bchapter}
\endbibitem

\bibitem{netfiter}
\begin{botherref}
Netfilter Architecture,
Accessed: 2019-02-20.
\end{botherref}
\endbibitem

\bibitem{Tracepoints}
\begin{botherref}
\oauthor{\binits{M.}~\bsnm{Desnoyers}},
Using the Linux Kernel Tracepoints,
Accessed: 2019-01-21.
\end{botherref}
\endbibitem

\bibitem{Huttonetal2003}
\begin{bchapter}
\bauthor{\binits{A.}~\bsnm{Hutton}},
\bauthor{\binits{T.}~\bsnm{Zanussi}},
\bauthor{\binits{K.}~\bsnm{Yaghmour}},
\bauthor{\binits{R.W.}~\bsnm{Wisniewski}},
\bauthor{\binits{R.}~\bsnm{Moore}} and
\bauthor{\binits{M.}~\bsnm{Dagenais}},
\bctitle{relayfs: An Efficient Unified Approach for Transmitting Data from
  Kernel to User Space},
in: \bbtitle{Proceedings of the Linux Symposium},
\bconflocation{Ottawa, Ontario, Canada}, \byear{2003}.
\url{https://www.kernel.org/doc/ols/2003/ols2003-pages-494-506.pdf}.
\end{bchapter}
\endbibitem

\bibitem{Prov_XML}
\begin{botherref}
\oauthor{\binits{L.}~\bsnm{Moreau}},
PROV-XML: The PROV XML Schema,
2013,
Accessed: 2019-06-12.
\end{botherref}
\endbibitem

\bibitem{Prov_JSON}
\begin{botherref}
\oauthor{\binits{T.D.}~\bsnm{Huynh}},
\oauthor{\binits{M.O.}~\bsnm{Jewell}},
\oauthor{\binits{A.S.}~\bsnm{Keshavarz}},
\oauthor{\binits{D.T.}~\bsnm{Michaelides}},
\oauthor{\binits{H.}~\bsnm{Yang}} and
\oauthor{\binits{L.}~\bsnm{Moreau}},
The PROV-JSON Serialization,
2013,
Accessed: 2019-01-05.
\end{botherref}
\endbibitem

\bibitem{Prov_Ontology}
\begin{botherref}
\oauthor{\binits{K.}~\bsnm{Belhajjame}},
\oauthor{\binits{J.}~\bsnm{Cheney}},
\oauthor{\binits{D.}~\bsnm{Corsar}},
\oauthor{\binits{D.}~\bsnm{Garijo}},
\oauthor{\binits{S.}~\bsnm{Soiland-Reyes}},
\oauthor{\binits{S.}~\bsnm{Zednik}} and
\oauthor{\binits{J.}~\bsnm{Zhao}},
PROV-O: The PROV Ontology,
2013,
Accessed: 2019-06-12.
\end{botherref}
\endbibitem

\bibitem{CASEY201714}
\begin{barticle}
\bauthor{\binits{E.}~\bsnm{Casey}},
\bauthor{\binits{S.}~\bsnm{Barnum}},
\bauthor{\binits{R.}~\bsnm{Griffith}},
\bauthor{\binits{J.}~\bsnm{Snyder}},
\bauthor{\binits{H.}~\bsnm{{van Beek}}} and
\bauthor{\binits{A.}~\bsnm{Nelson}},
\batitle{Advancing coordinated cyber-investigations and tool interoperability
  using a community developed specification language},
\bjtitle{Digital Investigation}
\bvolume{22}
(\byear{2017}),
\bfpage{14}--\blpage{45}.
doi:\doiurl{https://doi.org/10.1016/j.diin.2017.08.002}.
\url{http://www.sciencedirect.com/science/article/pii/S1742287617301007}.
\end{barticle}
\endbibitem

\bibitem{Pasquieretal2018b}
\begin{barticle}
\bauthor{\binits{T.}~\bsnm{Pasquier}},
\bauthor{\binits{J.}~\bsnm{Singh}},
\bauthor{\binits{J.}~\bsnm{Powles}},
\bauthor{\binits{D.}~\bsnm{Eyers}},
\bauthor{\binits{M.}~\bsnm{Seltzer}} and
\bauthor{\binits{J.}~\bsnm{Bacon}},
\batitle{Data provenance to audit compliance with privacy policy in the
  Internet of Things},
\bjtitle{Personal and Ubiquitous Computing}
\bvolume{22}(\bissue{2})
(\byear{2018}),
\bfpage{333}--\blpage{344}.
doi:\doiurl{10.1007/s00779-017-1067-4}.
\end{barticle}
\endbibitem

\bibitem{Braunetal2006}
\begin{bchapter}
\bauthor{\binits{U.}~\bsnm{Braun}},
\bauthor{\binits{S.}~\bsnm{Garfinkel}},
\bauthor{\binits{D.A.}~\bsnm{Holland}},
\bauthor{\binits{K.-K.}~\bsnm{Muniswamy-Reddy}} and
\bauthor{\binits{M.I.}~\bsnm{Seltzer}},
\bctitle{Issues in Automatic Provenance Collection},
in: \bbtitle{Provenance and Annotation of Data},
\beditor{\binits{L.}~\bsnm{Moreau}} and
\beditor{\binits{I.}~\bsnm{Foster}}, eds,
\bpublisher{Springer Berlin Heidelberg},
\blocation{Berlin, Heidelberg},
\byear{2006},
pp.~\bfpage{171}--\blpage{183}.
ISBN \bisbn{978-3-540-46303-0}.
\end{bchapter}
\endbibitem

\bibitem{Muniswamy-Reddy2006}
\begin{bchapter}
\bauthor{\binits{K.-K.}~\bsnm{Muniswamy-Reddy}},
\bauthor{\binits{D.A.}~\bsnm{Holland}},
\bauthor{\binits{U.}~\bsnm{Braun}} and
\bauthor{\binits{M.}~\bsnm{Seltzer}},
\bctitle{Provenance-aware Storage Systems},
in: \bbtitle{Proceedings of the Annual Conference on USENIX '06 Annual
  Technical Conference},
\bsertitle{ATEC '06},
\bpublisher{USENIX Association},
\blocation{Berkeley, CA, USA},
\byear{2006},
pp.~\bfpage{4}--\blpage{4}.
\url{http://dl.acm.org/citation.cfm?id=1267359.1267363}.
\end{bchapter}
\endbibitem

\bibitem{GehaniTariq2012}
\begin{bchapter}
\bauthor{\binits{A.}~\bsnm{Gehani}} and
\bauthor{\binits{D.}~\bsnm{Tariq}},
\bctitle{SPADE: Support for Provenance Auditing in Distributed Environments},
in: \bbtitle{Proceedings of the 13th International Middleware Conference},
\bsertitle{Middleware '12},
\bpublisher{Springer-Verlag New York, Inc.},
\blocation{New York, NY, USA},
\byear{2012},
pp.~\bfpage{101}--\blpage{120}.
ISBN \bisbn{978-3-642-35169-3}.
\url{http://dl.acm.org/citation.cfm?id=2442626.2442634}.
\end{bchapter}
\endbibitem

\bibitem{Leeetal2013a}
\begin{bchapter}
\bauthor{\binits{K.H.}~\bsnm{Lee}},
\bauthor{\binits{X.}~\bsnm{Zhang}} and
\bauthor{\binits{D.}~\bsnm{Xu}},
\bctitle{High Accuracy Attack Provenance via Binary-based Execution Partition},
in: \bbtitle{20th Annual Network and Distributed System Security Symposium,
  {NDSS} 2013, San Diego, California, USA, February 24-27, 2013},
\byear{2013}.
\url{https://www.ndss-symposium.org/ndss2013/high-accuracy-attack-provenance-binary-based-execution-partition}.
\end{bchapter}
\endbibitem

\bibitem{gnu_bison}
\begin{botherref}
GNU Bison,
Accessed: 2019-04-11.
\end{botherref}
\endbibitem

\bibitem{LLVM}
\begin{botherref}
The LLVM Compiler Infrastructure,
Accessed: 2019-04-25.
\end{botherref}
\endbibitem

\bibitem{Tanetal2018}
\begin{botherref}
\oauthor{\binits{C.}~\bsnm{Tan}},
\oauthor{\binits{Q.}~\bsnm{Wang}},
\oauthor{\binits{L.}~\bsnm{Wang}} and
\oauthor{\binits{L.}~\bsnm{Zhao}},
Attack Provenance Tracing in Cyberspace: Solutions, Challenges and Future
  Directions,
\textit{IEEE Network}
(2018),
1--7.
doi:\doiurl{10.1109/MNET.2018.1700469}.
\end{botherref}
\endbibitem

\bibitem{celik2019a}
\begin{botherref}
\oauthor{\binits{Z.B.}~\bsnm{Celik}},
\oauthor{\binits{E.}~\bsnm{Fernandes}},
\oauthor{\binits{E.}~\bsnm{Pauley}},
\oauthor{\binits{G.}~\bsnm{Tan}} and
\oauthor{\binits{P.}~\bsnm{McDaniel}},
Program Analysis of Commodity IoT Applications for Security and Privacy:
  Challenges and Opportunities,
\textit{ACM Comput. Surv.}
\textbf{52}(4)
(2019).
doi:\doiurl{10.1145/3333501}.
\end{botherref}
\endbibitem

\bibitem{Fernandes2016}
\begin{botherref}
\oauthor{\binits{E.}~\bsnm{Fernandes}},
\oauthor{\binits{J.}~\bsnm{Jung}} and
\oauthor{\binits{A.}~\bsnm{Prakash}},
{Security Analysis of Emerging Smart Home Applications},
\textit{Proceedings - 2016 IEEE Symposium on Security and Privacy, SP 2016}
(2016),
636--654.
ISBN 9781509008247.
doi:\doiurl{10.1109/SP.2016.44}.
\end{botherref}
\endbibitem

\bibitem{ftc2015}
\begin{botherref}
Internet of Things: Privacy and Security in a Connected World,
Federal Trade Commission,
Accessed: 2020-11-17.
\end{botherref}
\endbibitem

\bibitem{Elkhodr2019}
\begin{barticle}
\bauthor{\binits{M.}~\bsnm{Elkhodr}} and
\bauthor{\binits{Z.B.}~\bsnm{Mufti}},
\batitle{{On the Challenges of Data Provenance in The Internet of Things}},
\bjtitle{International Journal of Wireless {\&} Mobile Networks}
\bvolume{11}(\bissue{3})
(\byear{2019}),
\bfpage{43}--\blpage{52}.
doi:\doiurl{10.5121/ijwmn.2019.11304}.
\end{barticle}
\endbibitem

\bibitem{IFTTT2020}
\begin{botherref}
IFTTT: Every thing works better together,
Accessed: 2020-08-21.
\end{botherref}
\endbibitem

\bibitem{Celik2019b}
\begin{botherref}
\oauthor{\binits{Z.B.}~\bsnm{Celik}},
\oauthor{\binits{G.}~\bsnm{Tan}} and
\oauthor{\binits{P.}~\bsnm{Mcdaniel}},
{IOTGUARD : Dynamic Enforcement of Security and Safety Policy in Commodity IoT}
(2019).
ISBN 189156255X.
doi:\doiurl{10.14722/ndss.2019.23326}.
\end{botherref}
\endbibitem

\bibitem{Wangetal2018}
\begin{bchapter}
\bauthor{\binits{Q.}~\bsnm{Wang}},
\bauthor{\binits{W.}~\bsnm{{Ul Hassan}}},
\bauthor{\binits{A.}~\bsnm{Bates}} and
\bauthor{\binits{C.}~\bsnm{Gunter}},
\bctitle{{Fear and Logging in the Internet of Things}},
in: \bbtitle{Network and Distributed Systems Security Symposium},
\byear{2018}.
ISBN \bisbn{1-891562-49-5}.
doi:\doiurl{10.14722/ndss.2018.23282}.
\url{https://www.ndss-symposium.org/wp-content/uploads/2018/02/ndss2018_01A-2_Wang_paper.pdf}.
\end{bchapter}
\endbibitem

\bibitem{Prov_Overview}
\begin{botherref}
PROV-Overview,
Accessed: 2019-01-05.
\end{botherref}
\endbibitem

\bibitem{Hossainetal2017}
\begin{botherref}
\oauthor{\binits{N.}~\bsnm{Hossain}},
\oauthor{\binits{S.M.}~\bsnm{Milajerdi}},
\oauthor{\binits{J.}~\bsnm{Wang}},
\oauthor{\binits{B.}~\bsnm{Eshete}},
\oauthor{\binits{R.}~\bsnm{Gjomemo}},
\oauthor{\binits{R.}~\bsnm{Sekar}} and
\oauthor{\binits{S.}~\bsnm{Stoller}},
{SLEUTH : Real-time Attack Scenario Reconstruction from COTS Audit Data}
(2017),
487--504.
ISBN 9781931971409.
\end{botherref}
\endbibitem

\bibitem{Leonardo2018}
\begin{botherref}
\oauthor{\binits{L.}~\bsnm{Babun}},
\oauthor{\binits{A.K.}~\bsnm{Sikder}},
\oauthor{\binits{A.}~\bsnm{Acar}} and
\oauthor{\binits{A.S.}~\bsnm{Uluagac}},
IoTDots: {A} Digital Forensics Framework for Smart Environments,
\textit{CoRR}
\textbf{abs/1809.00745}
(2018).
\url{http://arxiv.org/abs/1809.00745}.
\end{botherref}
\endbibitem

\bibitem{Liu2019}
\begin{barticle}
\bauthor{\binits{Z.}~\bsnm{Liu}} and
\bauthor{\binits{Y.}~\bsnm{Wu}},
\batitle{{An Index-based Provenance Compression Scheme for Identifying
  Malicious Nodes in Multi-hop IoT Network}},
\bjtitle{IEEE Internet of Things Journal}
\bvolume{14}(\bissue{8})
(\byear{2019}),
\bfpage{1}--\blpage{1}.
doi:\doiurl{10.1109/jiot.2019.2961431}.
\end{barticle}
\endbibitem

\bibitem{Zheng2019}
\begin{botherref}
\oauthor{\binits{R.}~\bsnm{Zheng}},
\oauthor{\binits{J.}~\bsnm{Jiang}},
\oauthor{\binits{X.}~\bsnm{Hao}},
\oauthor{\binits{W.}~\bsnm{Ren}},
\oauthor{\binits{F.}~\bsnm{Xiong}} and
\oauthor{\binits{Y.}~\bsnm{Ren}},
{BcBIM: A Blockchain-Based Big Data Model for BIM Modification Audit and
  Provenance in Mobile Cloud},
\textit{Mathematical Problems in Engineering}
\textbf{2019}
(2019).
doi:\doiurl{10.1155/2019/5349538}.
\end{botherref}
\endbibitem

\bibitem{Suhail2020}
\begin{barticle}
\bauthor{\binits{S.}~\bsnm{Suhail}},
\bauthor{\binits{R.}~\bsnm{Hussain}},
\bauthor{\binits{M.}~\bsnm{Abdellatif}},
\bauthor{\binits{S.R.}~\bsnm{Pandey}},
\bauthor{\binits{A.}~\bsnm{Khan}} and
\bauthor{\binits{C.S.}~\bsnm{Hong}},
\batitle{{Provenance-enabled packet path tracing in the RPL-based internet of
  things}},
\bjtitle{Computer Networks}
\bvolume{173}
(\byear{2020}),
\bfpage{1}--\blpage{15}.
doi:\doiurl{10.1016/j.comnet.2020.107189}.
\end{barticle}
\endbibitem

\bibitem{Kamal2019}
\begin{barticle}
\bauthor{\binits{M.}~\bsnm{Kamal}} and
\bauthor{\binits{M.}~\bsnm{Tariq}},
\batitle{{Light-weight security and blockchain based provenance for advanced
  metering infrastructure}},
\bjtitle{IEEE Access}
\bvolume{7}
(\byear{2019}),
\bfpage{87345}--\blpage{87356}.
doi:\doiurl{10.1109/ACCESS.2019.2925787}.
\end{barticle}
\endbibitem

\end{thebibliography}

%

\end{document}